\documentclass[12pt]{article}
\usepackage{amsmath}
\usepackage[dvips]{graphicx}
\usepackage{epsfig}
\usepackage{amsmath}
\usepackage{amssymb}
\usepackage{graphics,epstopdf}
\usepackage{amsfonts}
\usepackage{amsthm}
\usepackage[usenames]{color}

\definecolor{AV}{rgb}{0.65,0.0,0}
\definecolor{GC}{rgb}{0,0.0,0.65}
\definecolor{WS}{rgb}{0,0.65,0}

\usepackage[
      colorlinks=true,
      linkcolor=blue,
      urlcolor=blue,
      filecolor=blue,
      citecolor=red,
      pdfstartview=FitV,
      pdftitle={},
      pdfauthor={},
      pdfsubject={},
      pdfkeywords={},
      pdfpagemode=None,
      bookmarksopen=true
]{hyperref}

\newcommand{\bm}{\begin{multiline}}
\newcommand{\beq}{\begin{equation}}
\newcommand{\eeq}{\end{equation}}
\newcommand{\beqs}{\begin{eqnarray}}
\newcommand{\eeqs}{\end{eqnarray}}

\begin{document}

\thispagestyle{empty}

\hfill{}

\hfill{}

\hfill{}

\vspace{32pt}

\begin{center}

\textbf{\Large New bound of the mass-to-radius ratio for electrically charged stars}

\vspace{48pt}

\textbf{ Cristian Stelea,}\footnote{E-mail: \texttt{cristian.stelea@uaic.ro}}
\textbf{Marina-Aura Dariescu,}\footnote{E-mail: \texttt{marina@uaic.ro}}
\textbf{Ciprian Dariescu, }\footnote{E-mail: \texttt{ciprian.dariescu@uaic.ro}}

\vspace*{0.2cm}

\textit{$^1$ Research Department, Faculty of Physics, ``Alexandru Ioan Cuza" University}\\[0pt]
\textit{11 Bd. Carol I, Iasi, 700506, Romania}\\[.5em]

\textit{$^{2,3}$ Faculty of Physics, ``Alexandru Ioan Cuza" University}\\[0pt]
\textit{11 Bd. Carol I, Iasi, 700506, Romania}\\[.5em]

\end{center}

\vspace{30pt}

\begin{abstract}
Using a general solution-generating technique for electrically charged relativistic stars with spherical symmetry, we derive a new bound on the mass-radius ratio. This compactness bound is based on the already established bounds for uncharged interior solutions and it will provide the corresponding generalizations in presence of the electric charge.
\end{abstract}

\vspace{32pt}

\setcounter{footnote}{0}

\newpage

\section{Introduction}

According to the Birkhoff theorem the most general description of the outside geometry of a nonrotating and spherically symmetric relativistic star is provided by the Schwarzschild solution. There is however some freedom in choosing the specific interior solution for the compact star. Ever since the beginning of the General Relativity (GR) as a theory of gravity and the pioneering works of Schwarzschild \cite{Schwarzschild:1916uq} and Tolman \cite{Tolman:1939jz} compact objects were usually modeled using spherically symmetric perfect fluid solutions of the Einstein equations. There are known by now various solution-generating techniques that allow us to find new spherically symmetric solutions of the Einstein equations sourced by perfect fluids \cite{Lake:2002bq} - \cite{Boonserm:2005ni}. One should also note at this point that not all the perfect fluid solutions that can be generated this way are also physical \cite{Delgaty:1998uy}. 

On the other hand, spherical symmetry also allows more general anisotropic fluid configurations of the star's interior. For such fluid configurations the anisotropy means that the pressure along the radial direction $p_r$ is usually different from the pressures along the transverse directions, $p_t$. There are many reasons to consider anisotropic fluids as interior models for relativistic stars \cite{Ruderman:1972aj}. The anisotropy of the interior fluid distribution can arise from various reasons: it can be due to a mixture of two fluid components \cite{Letelier:1980mxb}, the existence of a superfluid phase, the presence of a magnetic field, etc. (for a review see \cite{Herrera:1997plx} and references there). There are also known other non-trivial examples of anisotropic fluid distributions, such as the bosonic stars (see for instance \cite{Liebling:2012fv} and the references within), or traversable wormholes \cite{Bronnikov:2017kvq}, or the so-called gravastars \cite{Cattoen:2005he}, which are systems where anisotropic pressures occur naturally. Not surprisingly, in the last decades there has been renewed interest in deriving new physical solutions with interior anisotropic fluids (see for instance \cite{Harko:2002db} - \cite{Herrera:2007kz}).

In general, for a spherically symmetric compact object the compactness is defined using the mass-to-radius ratio, $\frac{M}{R}$. The compactness is essential in the determination of the outside geometry of the star and it is also a measure of the strength of the gravitational field of that compact object. One can also perform direct experimental observations of the compactness of a relativistic star since the compactness is also related to the gravitational redshift of that object (see for instance \cite{Sieniawska:2018pev}, \cite{Ozel:2016oaf} and the references therein). Since the seminal work of Buchdahl \cite{Buchdahl} it is well-known that a spherically compact object whose energy density decreases monotonically will present an upper bound on the mass-to-radius ratio:
\beqs
\frac{M}{R}&\leq&\frac{4}{9}.
\label{b1}
\eeqs
Here $M$ is the ADM mass of the compact object and $R$ is its radius. The critical value $\frac{M}{R}=\frac{4}{9}$ corresponds to the limit case of the interior Schwarzschild solution \cite{Schwarzschild:1916uq} for which the pressure at the center of the star becomes infinite. However, it was shown in \cite{Andreasson:2007ck}, \cite{Karageorgis:2007cy} that this bound also holds in more general anisotropic configurations that satisfy the energy condition:
\beqs
p_r+2p_t\leq \rho,
\eeqs
where $\rho$ is the fluid's energy density. Moreover, for a more general energy condition of the form $p_r+2p_t\leq \Omega\rho$ with $\Omega>0$, it was shown in \cite{Andreasson:2007ck} that the mass-to radius ratio is bounded by:
\beqs
\frac{M}{R}&\leq&\frac{1}{2}\frac{(1+2\Omega)^2-1}{(1+2\Omega)^2},
\label{bo}
\eeqs
where $\rho$ and $p$ are nonnegative. For $\Omega=1$ one obviously recovers the bound (\ref{b1}).

Bounds for spherically-symmetric static solutions of the Einstein-fluid equations in presence of a positive cosmological constant have been found in \cite{Andreasson:2009pe}, \cite{Boehmer:2006ye}. For configurations that describe non-compact objects the discussion is a bit more involved since these configurations do not have a sharp boundary on which the radial pressure vanishes. However, even in these cases Buchdahl-type of inequalities have been derived in \cite{Hod:2007sy}.
 
 For charged configurations the similar bounds have been established in \cite{Andreasson:2008xw}, \cite{Mak:2001ie}. Of particular importance, due to its simplicity is the bound derived in \cite{Andreasson:2008xw}:
 \beqs
 \frac{M}{R}&\leq&\frac{2}{9}+\frac{Q^2}{3R^2}+\sqrt{1+\frac{3Q^2}{R^2}}.
 \label{ba}
 \eeqs
 This bound has been further generalized in \cite{Andreasson:2012dj} to allow for the presence of a cosmological constant.
 
 In this paper we  will derive a similar simple bound for electrically charged objects, solutions of the full Einstein-Maxwell-fluid model. Our bound will be based on the previously derived bounds for uncharged objects as given in (\ref{b1}) or (\ref{bo}). From the uncharged interior solution we shall construct the corresponding electrically charged solution. We stress that this approach is based on an exact solution of the Einstein-Maxwell-fluid equations of motion.
One such simple solution of the electrically charged interior solution of the full Einstein-Maxwell-fluid theory, for more general geometries with axial symmetry has been provided in \cite{Stelea:2018elx}. In our work we shall adapt this method to the spherically symmetric case and also use a slightly more general form than that provided in \cite{Stelea:2018elx}. The more general solution contains an extra parameter that we shall judiciously choose in order to obtain asymptotically flat exterior solutions. By computing the mass and charge of the outside geometry of the star, which is the Reissner-Nordstr\"{o}m solution, our Buchdahl-type bound for charged objects takes the form:
\beqs
\frac{M}{R}&\leq&\frac{4}{9}\bigg[5\sqrt{1+\frac{9Q^2}{16R^2}}-4\bigg].
\label{our}
\eeqs

The structure of this paper is as follows: in the next section we present the general electrically charged version of a spherically symmetric interior solution. In section $3$ we show how to derive the charged bound (\ref{our}) by using the bound (\ref{b1}) for  the uncharged solution of the Einstein equations. The final section contains a summary of our work and avenues for further work.

\section{The electrically charged solution}

Our starting point is the solution-generating technique presented in \cite{Stelea:2018elx}. We will adapt here their results for a spherically symmetric spacetime with the line element:
\beqs
ds^2&=&-g_{tt}(r)dt^2+g_{rr}(r)dr^2+r^2(d\theta^2+\sin^2\theta d\varphi^2).
\label{uncharged}
\eeqs
In general, this geometry is a solution of the Einstein-fluid equations:
\beqs 
G_{\mu\nu}&=&8\pi T_{\mu\nu}^0,
\eeqs 
where the stress-energy $T_{\mu\nu}^0$ of the anisotropic distribution of matter has the form:
\beqs
T_{\mu\nu}^0&=&(\rho^0+p_t^0)u_{\mu}^0u_{\nu}^0+p_t^0g_{\mu\nu}^0+(p_r^0-p_t^0)\chi_{\mu}^0\chi_{\nu}^0.
\eeqs
Here $\rho^0$ is the fluid density, $p_r^0$ is the radial component of the pressure, while $p_t^0$ represents the transverse components of the pressure. Moreover, $u_{\mu}^0$ is the $4$-velocity of the fluid while $(\chi^0)^{\mu}=\sqrt{g_{rr}^{-1}}\delta^{\mu}_r$ is the unit spacelike vector in the radial direction. 

According to the results in \cite{Stelea:2018elx}, the corresponding electrically charged solution is:
\beqs
ds^2&=&-\frac{g_{tt}(r)}{\Lambda^2}dt^2+\Lambda^2\big[g_{rr}(r)dr^2+r^2(d\theta^2+\sin^2\theta d\varphi^2)\big],
\label{chargedm}
\eeqs
where we defined $\Lambda=\frac{1-U^2g_{tt}(r)}{C}$. This is an exact solution of the Einstein-Maxwell-fluid equation:
\beqs
G_{\mu\nu}&=&8\pi T^{EM}_{\mu\nu}+8\pi T^{af}_{\mu\nu},~~~~F^{\mu\nu}_{;\nu}=4\pi J^{\mu}.\label{EMAI}
\eeqs
where the stress-energy tensor of the anisotropic fluid is given by:
\beqs
T^{af}_{\mu\nu}&=&(\rho+p_t+\sigma_e)u_{\mu}u_{\nu}+p_tg_{\mu\nu}+(p_r-p_t)\chi_{\mu}\chi_{\nu}.
\eeqs
Here $\rho=\frac{\rho^0}{\Lambda^2}$ is the energy density of the fluid, $p_r=\frac{p_r^0}{\Lambda^2}$ is the radial pressure and $p_t=\frac{p_t^0}{\Lambda^2}$ is the transverse pressure of the fluid. The charge density of the fluid is given by:
\beqs
\sigma_e&=&\frac{2}{C}(\rho+p_r+2p_t)\frac{Ug_{tt}(r)}{\Lambda}.
\eeqs
Finally, the electromagnetic stress-energy tensor in (\ref{EMAI}) is defined as:
\beqs
T^{EM}_{\mu\nu}&=&\frac{1}{4\pi}\left(F_{\mu\alpha}F^{~\alpha}_{\nu}-\frac{1}{4}F^2g_{\mu\nu}\right).
\eeqs
Here the Maxwell field is described by the $4$-potential $A_{\mu}=(\frac{Ug_{tt}(r)}{\Lambda}, 0, 0, 0)$ and it is sourced by the electric $4$-current $J_{\mu}=(j_t, 0, 0, 0)$ where:
\beqs
j_t&=&-\frac{2}{C}(\rho+p_r+2p_t)\frac{Ug_{tt}(r)}{\Lambda^2}.
\eeqs

Note that the above solution contains two parameters $U$ and $C$ and it is slightly more general than the one presented in \cite{Stelea:2018elx}.

\section{The charged bound for the mass-to-radius ratio}

For a regular uncharged compact object the exterior geometry is that of the Schwarzschild vacuum solution:
\beqs
ds^2&=&-\left(1-\frac{2m}{r}\right)dt^2+\frac{dr^2}{1-\frac{2m}{r}}+r^2(d\theta^2+\sin^2\theta d\varphi^2),
\label{sch}
\eeqs
where $m$ is the mass of the uncharged configuration. The uncharged interior solution (\ref{uncharged}) will match the Schwarzschild exterior geometry (\ref{sch}) on the star's surface, which is defined by $p_r^0(r_0)=0$. In the electrically charged solution the exterior geometry corresponds to the charged version (\ref{chargedm}) of (\ref{sch}), \footnote{For the exterior solution the fluid density and pressures vanish.} which should be the Reissner-Nordstr\"{o}m solution. Indeed, in order to have an asymptotically flat exterior solution it is convenient at this point to pick the value of the constant $C=1-U^2$. It is now easy to see that by performing the coordinate transformation:
\beqs
r&=&R-\frac{2mU^2}{1-U^2}
\eeqs
the exterior geometry becomes the Reissner-Nordstr\"{o}m solution written in the canonical form:
\beqs
ds^2&=&-\left(1-\frac{2M}{R}+\frac{Q^2}{R^2}\right)dt^2+\frac{dR^2}{1-\frac{2M}{R}+\frac{Q^2}{R^2}}+R^2(d\theta^2+\sin^2\theta d\varphi^2),\\
A_t&=&U+\frac{Q}{R}\nonumber,
\label{rn}
\eeqs
where the ADM mass $M$ and charge $Q$ of the exterior solution are given by:
\beqs
M&=&\frac{1+U^2}{1-U^2}m,~~~Q=\frac{2mU}{1-U^2}.
\eeqs

This exterior geometry will match continuously the charged interior on the star surface $r=r_0$. Note that using the new radial coordinate $R$ the star surface is now located at $R_0=r_0+\frac{2mU^2}{1-U^2}$. Moreover, since $m=\sqrt{M^2-Q^2}$ one has $\frac{Q}{M}\leq 1$. Finally, the mass-to-radius ratio for the uncharged configuration can be cast into the form:
\beqs
\frac{m}{r_0}&=&\frac{\sqrt{M^2-Q^2}}{R_0-M+\sqrt{M^2-Q^2}}
\label{bound0}
\eeqs

Assuming now that the Buchdahl bound (\ref{b1}) is valid for the uncharged configuration, then using the above relation it is easy to derive the corresponding compactness bound in the electrically charged case:
\beqs
\frac{M}{R_0}&\leq&\frac{4}{9}\bigg[5\sqrt{1+\frac{9Q^2}{16R_0^2}}-4\bigg].
\eeqs
It should be obvious that in the uncharged case $Q=0$ the above inequality reduces to the Buchdahl bound (\ref{b1}). It is also possible to restate the above inequality from (\ref{bound0}) by using the extremality parameter $\frac{Q}{M}$ instead of $q=\frac{Q}{R_0}$. In this case one has:
\beqs
\frac{M}{R_0}&\leq&\frac{4}{4+5\sqrt{1-\frac{Q^2}{M^2}}}.
\eeqs
In the extremal case $\frac{Q}{M}=1$ and the bound becomes $\frac{M}{R_0}\leq 1$.

It would be interesting to compare the above charged bound (\ref{our}) with the previously known bound (\ref{ba}) found in \cite{Andreasson:2007ck}. Defining the ratio $q=\frac{Q}{R_0}$ we shall consider the difference between the two bounds:
\beqs
f(q)&=&\frac{4}{9}\bigg[5\sqrt{1+\frac{9q^2}{16}}-4\bigg]-\left(\frac{2}{9}+\frac{q^2}{3}+\sqrt{1+3q^2}\right)
\label{fq}
\eeqs

\begin{figure}
  \centering
  \includegraphics[width=0.45\textwidth]{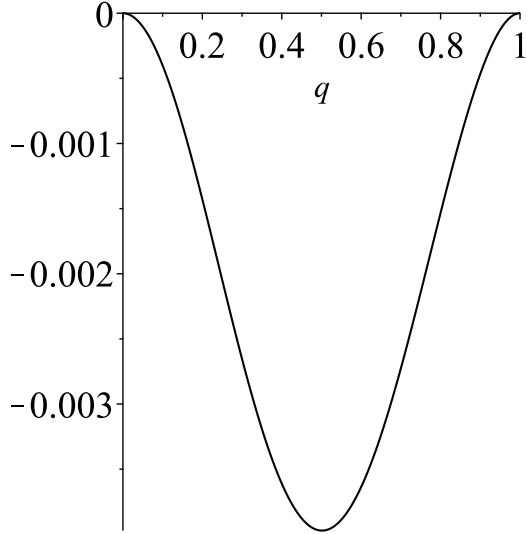} 
  \caption{Plot of the function $f(q)$ in (\ref{fq}) versus the parameter $q$.} 
  \label{figure}
\end{figure}

In Figure \ref{figure} we plot the difference (\ref{fq}) between the two inequalities for the physical values of the charge-to-radius parameter $q$ from $0$ to $1$. Note that this function is always negative in this interval and therefore our bound in (\ref{our}) limits the compactness values for charged objects even more than the bound in (\ref{ba}). However, one should note at this point that in our solution-generating technique the value of the $tt$-component of the metric in the charged solution (\ref{chargedm}) (let us denote it here by $(-G_{tt})$) is related to the electric potential in the charged solution by a Weyl-type relation of the form:
\beqs
G_{tt}&=&A_t^2+\frac{C}{U}A_t.
\label{weyl}
\eeqs
This relation between the metric component $G_{tt}$ and the electric potential $A_t$ is not the most general that one could envisage. For instance, there are known other exact solutions of the Einstein-Maxwell-fluid equations that generalize the above relation. One such example is provided by the Guilfoyle stars \cite{Guilfoyle:1999yb}. In \cite{Lemos:2015wfa} it was shown that the Andreasson bound (\ref{ba}) is saturated for a particular Guilfoyle solution for which $-G_{tt}=aA_t^2$. However, this solution is not covered by our solution-generating technique so we will not discuss it here.

One might wonder what is the charged configuration that could reach the upper bound in (\ref{our}). For instance, in the uncharged case the Buchdahl inequality is saturated by the interior Schwarzschild solution if the compactness is $\frac{m}{r_0}=\frac{4}{9}$. As it turns out, our inequality is saturated by the corresponding charged version of the interior Schwarzschild solution, as expected.

Finally, let us mention that by using (\ref{bound0}) and (\ref{bo}) one could easily generalize the bound (\ref{our}) for the case in which $\Omega\neq 1$. However, the result is not particularly illuminating and we will not list it here.

\section{Conclusions}

In this work we found a new bound for the mass-to-radius ratio for charged compact objects. Our bound is based on the well-known Buchdahl bound for uncharged objects. The novelty of our approach is that we used a solution-generating technique to find the corresponding charged solution. In this way we were able to find directly the relation between the compactness of the charged star and the compactness for the corresponding uncharged star. Our bound turns out to be smaller than the previous charged bound found by Andreasson in \cite{Andreasson:2007ck}. However, we found that the charged interior solution that saturates the bound is the charged interior Schwarzschild solution in the particular limit case in which the radial pressure blows up at origin, as expected.

As avenues for further work, it would be interesting to find extensions (if any) of our solution-generating technique to the more general cases that might include the Guilfoyle solutions (more general Weyl ansatzes in (\ref{weyl})). Also it might prove instructive to find similar bounds for electrically charged interior solutions in higher dimensions. Work on these issues is in progress and it will be reported elsewhere.

\vspace{10pt}

{\Large Acknowledgements}

This work was supported by a grant of Ministery of Research and Innovation, CNCS - UEFISCDI, project number PN-III-P4-ID-PCE-2016-0131, within PNCDI III.

\end{document}